\documentstyle[12pt]{article}
\begin{document}
\setlength{\baselineskip}{20pt}

\mbox{}

\begin{center}
{\Large{\bf New solutions of the Jacobi equations for three-dimensional Poisson structures}} \\ 
\mbox{} \\
{\bf Benito Hern\'{a}ndez--Bermejo$^{\; \mbox{\footnotesize {\rm a)}}}$}
\end{center}
\noindent {\em Departamento de F\'{\i}sica Matem\'{a}tica y Fluidos, Universidad Nacional de 
Educaci\'{o}n a Distancia. Senda del Rey S/N, 28040 Madrid, Spain.}

\mbox{}

\begin{center} 
{\bf Abstract}
\end{center}

A systematic investigation of the skew-symmetric solutions of the three-dimensional Jacobi 
equations is presented. As a result, three disjoint and complementary new families of solutions 
are characterized. Such families are very general, thus unifying 
many different and well-known Poisson structures seemingly unrelated which now appear embraced 
as particular cases of a more general solution. This unification is not only conceptual but 
allows the development of algorithms for the explicit determination of important properties 
such as the symplectic structure, the Casimir invariants and the Darboux canonical form, which 
are known only for a limited sample of Poisson structures. These common procedures are thus 
simultaneously valid for all the particular cases which can now be analyzed in a unified and 
more economic framework, instead of using a case-by-case approach. In addition, the methods 
developed are valid globally in phase space, thus ameliorating the usual scope of Darboux' 
reduction which is only of local nature. Finally, the families of solutions found present some 
new nonlinear superposition principles which are characterized.

\mbox{}

\mbox{}



\noindent {\bf Short Title:} New solutions of the 3D Jacobi equations.

\mbox{}

\mbox{}

\mbox{}

\footnoterule
\noindent $^{\mbox{\footnotesize {\rm a)}}}$ {\bf Present address:} 
Universit\'{e} Libre de Bruxelles (ULB), Service de Physique Th\'{e}orique et Math\'{e}matique, 
Campus Plaine -- CP 231, Boulevard du Triomphe, B-1050 Bruxelles, Belgium. 
\newline E-mail: bhernan@cso.ulb.ac.be 

\pagebreak
\begin{flushleft}
{\bf I. INTRODUCTION}
\end{flushleft}

Poisson structures$^{1,2 \:}$ have an important presence in all fields of Mathematical 
Physics.$^{1-35 \:}$  A Poisson description of a given system is often the basis for the 
obtainment of fruitful insight and information through the use of a plethora of well-known 
adapted tools.$^{1,2,7,9,17,34,36-42 \:}$

The present work is devoted to finite-dimensional Poisson structures. These, when expressed in 
terms of a system of local coordinates on an $n$-dimensional manifold, take the form:
\begin{equation}
    \label{nham}
    \dot{x}_i = \sum_{j=1}^n J_{ij} \partial _j H \; , \;\:\; i = 1, \ldots , n
\end{equation} 
Here and in what is to follow $ \partial_j $ means $ \partial / \partial x_j$.
The $C^1$ and real-valued function $H(x)$ in (\ref{nham}) is a constant of motion of the 
system, which plays the role of Hamiltonian. The $J_{ij}(x)$, called structure functions, are 
also $C^1$ and real-valued and constitute the entries of a $n \times n$ structure matrix 
${\cal J}$. The $J_{ij}(x)$ have the property of being solutions of the Jacobi equations:
\begin{equation}
     \label{jac}
     \sum_{l=1}^n ( J_{li} \partial_l J_{jk} + J_{lj} \partial_l J_{ki} + 
     J_{lk} \partial_l J_{ij} ) = 0 
\end{equation}
In (\ref{jac}) indices $i,j,k$ run from 1 to $n$. The structure functions also verify the additional 
condition of being skew-symmetric:
\begin{equation}
      \label{sksym}
	J_{ij}=-J_{ji} \;\:\;\: \mbox{for all} \; \: i,j 
\end{equation}

One of the reasons justifying the importance of the Poisson representation is the local 
equivalence bewteen Poisson systems and classical Hamiltonian systems, as stated by Darboux 
Theorem:$^{1 \:}$

\mbox{}

{\bf Theorem 1.1 (Darboux): \/} Consider an $n$-dimensional Poisson manifold for which the 
rank of the Poisson structure has constant value $2r$ everywhere. Then at each point of the 
manifold there exist local coordinates $(p_1, \ldots ,p_{r},q_1, \ldots , q_{r},z_1, \ldots , 
z_{n-2r})$ in terms of which the equations of motion become:
\[
	\dot{q}_i = \frac{\partial H}{\partial p_i} \;\: , \;\:\:
	\dot{p}_i = - \frac{\partial H}{\partial q_i} \;\: , \;\:\: i=1, \ldots ,r 
\]
\[
	\dot{z}_j = 0 \;\: , \;\:\: j=1 , \ldots , n-2r
\]

\mbox{}

An interesting consequence of Darboux Theorem which will be reconsidered later is that two 
Poisson structures of the same dimension and rank can be transformed locally into each other by 
a suitable change of coordinates.$^{1 \:}$ Both the Darboux Theorem and the previous remark are 
important for what follows.

The possibility of describing a given vector field not explicitly written in the form 
(\ref{nham}) in terms of a Poisson structure is an obvious question of fundamental importance 
in this context to which important efforts have been devoted in past years in a variety of 
approaches.$^{5-17,19,21-28 \:}$  This explains, together with the intrinsic 
mathematical interest of the problem, the permanent attention deserved in the literature by 
the obtainment and classification of skew-symmetric solutions of the Jacobi 
equations,$^{3-18,20-24,26,27,29 \:}$ both in the case of $n$-dimensional 
solu\-tions$^{3,4,13,14,18,20,26,29 \:}$ as well as in the important and better understood 
situation of dimension three.$^{7-12,17,22,23,27 \:}$ In the three-dimensional case the 
strategy for finding suitable skew-symmetric solutions of the Jacobi equations makes use of 
very diverse ---often problem-dependent--- methods, and the state of affairs is certainly more 
elaborate than the one existing in general dimension $n$. Moreover, it is worth 
recalling that the three-dimensional scenario is particularly relevant for several reasons: 
First, a large number of three-dimensional systems 
arising in very diverse fields have a Poisson structure (see Tables I to IV for a sample). 
Therefore three-dimensional Poisson structures are the natural framework for their analysis. 
Second, dimension three corresponds to the first nontrivial case where Poisson 
structure does not imply symplectic structure, i.e. it is the simplest meaningful kind of 
Poisson structures which is not symplectic. And third, three is the lowest dimension for which 
the Jacobi identities are not always identically verified. Since the complexity of equations 
(\ref{jac}-\ref{sksym}) is increasing with the dimension $n$, the three-dimensional case is the 
simplest nontrivial one as well as a natural first approach to the full problem of analyzing 
system (\ref{jac}-\ref{sksym}).

In this work a systematic investigation of the skew-symmetric solutions of the three-dimensio\-nal 
Jacobi equations (\ref{jac}-\ref{sksym}) is presented. As we shall see, three disjoint 
categories of solutions of the problem appear naturally. For each of them, a new family of 
solutions is found. Such families are extremely general. This explains that many well-known 
three-dimensional Poisson structures and dynamical systems now happen to appear embraced as 
particular cases of a wider family, as we shall see in detail. Therefore, a first outcome is 
that of the unification of many different Poisson structures seemingly unrelated. Moreover, 
this unification is not only conceptual. In fact, the new families are amenable to explicit 
and detailed analysis, in spite of their generality. In particular, it is possible to develop 
algorithms for the determination of important properties such as the symplectic structure and 
the Darboux canonical form. The advantage of these common strategies is that they are 
simultaneously valid for all the particular cases which can now be analyzed in a unified and 
very economic way, instead of using a case-by-case approach. In addition, the methods developed 
are valid globally in phase space, thus ameliorating the usual scope of Darboux' theorem which 
does only guarantee, in principle, a local reduction.$^{1 \:}$ The possibility of constructing 
the Darboux canonical form is also remarkable in view that the practical determination of 
Darboux' coordinates is a complicated task in general, which has been carried out only for a 
very limited sample of systems.$^{2,6,13,14,19 \:}$ Finally, the families of solutions found 
have unexpected properties, such as the presence of simple nonlinear superposition principles 
which will be characterized.

For the sake of conciseness, in what follows we shall use the following notation for the 
entries of the three-dimensional structure matrix:
\begin{equation}
	\label{uvw}
	u(x) := J_{12}(x) \; , \;\:\; v(x) := J_{31}(x) \; , \;\:\; w(x) := J_{23}(x)
\end{equation}
Now, if in the case $n=3$ we simplify the Jacobi identities (\ref{jac}) with the help of 
(\ref{sksym}) and substitute also definition (\ref{uvw}) the joint system 
(\ref{jac}-\ref{sksym}) takes the form:
\begin{equation}
     \label{jac3df}
     u \partial_1 v - v \partial_1 u + 
     w \partial_2 u - u \partial_2 w + v \partial_3 w - w \partial_3 v = 0 
\end{equation}
The three-dimensional version of system (\ref{jac}-\ref{sksym}) shall be written in the compact 
form (\ref{jac3df}) in the rest of this work.

The structure of the article is as follows. In Sections II to IV, respectively, three 
different, disjoint and complementary families of solutions are developed including their 
derivation and properties as well as examples. To conclude, Section V contains some final 
remarks.

\mbox{}

\begin{flushleft}
{\bf II. FIRST FAMILY OF SOLUTIONS}
\end{flushleft}

For the characterization of the first family of solutions, it is convenient to begin with the
establishment of an important general property of equation (\ref{jac3df}):

\mbox{}

{\bf Theorem 2.1: \/} Let $\{u(x),v(x),w(x)\}$ be a set of $C^1(\Omega)$ functions solution of 
equation (\ref{jac3df}) in an open domain $\Omega \subset I \!\! R^3$, and let $\mu(x) : \Omega 
\longrightarrow I \!\! R$ be an arbitrary $C^1(\Omega)$ function. Then $\{u^*(x),v^*(x),w^*(x)\}
= \{\mu(x) u(x),\mu(x) v(x),\mu(x) w(x)\}$ is also a solution of equation (\ref{jac3df}).

\mbox{}

{\em Proof: \/} It can be verified after direct substitution of $\{\mu(x) u(x),\mu(x) v(x),
\mu(x)w(x)\}$ into equation (\ref{jac3df}). \hfill Q.E.D.

\mbox{}

It is important to stress that this theorem is not valid in general in dimensions higher than 
three, as it can be easily verified. In order to physically interpret the result contained in 
Theorem 2.1 it is necessary to first formalize the concept of time reparametrization:$^{6 \:}$

\mbox{}

{\em Definition 2.2: \/} Let $\Omega \subset I \!\! R^3$ be an open subset. A reparametrization 
of time is defined as a transformation of the form
\begin{equation}
	\label{ntt}
	\mbox{d}\tau = \frac{1}{\mu(x)}\mbox{d}t
\end{equation}
where $t$ is the initial time variable, $\tau$ is the new time and $\mu(x) : \Omega 
\longrightarrow I \!\! R$ is a $C^1(\Omega)$ function which does not vanish in $\Omega$.

\mbox{}

In addition, let
\begin{equation}
	\label{3dpos}
	\frac{\mbox{d}x}{\mbox{d}t} = {\cal J} \cdot \nabla H
\end{equation}
be an arbitrary three-dimensional Poisson structure defined in an open domain 
$\Omega \subset I \!\! R^3$. Then, every reparametrization of time of the form (\ref{ntt}) 
leads from (\ref{3dpos}) to the differential system:
\begin{equation}
	\label{3dposntt}
	\frac{\mbox{d}x}{\mbox{d} \tau} = \mu {\cal J} \cdot \nabla H
\end{equation}
Consequently, in the three-dimensional case reparametrizations (\ref{ntt}) preserve the existence 
of a Poisson structure in the system, this time with structure matrix $\mu {\cal J}$ in 
(\ref{3dposntt}). On the contrary, such transformations in general destroy the Poisson 
structure in higher dimensions because for a given ${\cal J}$ which is a structure matrix, 
$\mu {\cal J}$ is not necessarily a solution of (\ref{jac}-\ref{sksym}).

We proceed now to characterize a first family of solutions of equation (\ref{jac3df}). For 
this, we shall assume that none of the solution functions $\{u(x),v(x),w(x)\}$ is identically 
zero (the relaxation of this condition will lead to the other two families of solutions, as 
we shall see in Sections III and IV). 

\mbox{}

{\em Definition 2.3: \/} For every open domain $\Omega \subset I \!\! R^3$, we shall denote by 
$\Gamma_{[u,v,w]}(\Omega)$ the set of solutions of equation (\ref{jac3df}) defined in $\Omega$ 
which are of the form $\{ u(x),v(x),w(x) \}$, with $u(x)$, $v(x)$ and $w(x)$ nonvanishing in 
$\Omega$ and $C^1(\Omega)$.

\mbox{}

We have the following result:

\mbox{}

{\bf Theorem 2.4: \/} Consider the family of functions of the form
\begin{equation}
	\label{sol1expl}
	\left\{ \begin{array}{ccl} 
	u(x) & = & \eta (x) \psi _1(x_1) \psi _2(x_2) \phi _3(x_3)    \\
	v(x) & = & \eta (x) \psi _1(x_1) \phi _2(x_2) \psi _3(x_3)    \\
	w(x) & = & \eta (x) \phi _1(x_1) \psi _2(x_2) \psi _3(x_3) 
	\end{array} \right.
\end{equation}
defined in an open set $\Omega \subset I \!\! R^3$, where $\{ \eta , \psi _i, \phi _i \}$, 
$i=1,2,3$, are arbitrary $C^1(\Omega)$ functions of their respective arguments which do not 
vanish in $\Omega$. Then the family of functions (\ref{sol1expl}) belongs to 
$\Gamma _{[u,v,w]}(\Omega)$.

\mbox{}

{\em Proof: \/} For solutions belonging to $\Gamma _{[u,v,w]}(\Omega)$, we can equivalently 
write (\ref{jac3df}) as:
\begin{equation}
     \label{jac3dc1}
      u^2 \partial_1 \left( \frac{v}{u} \right) + 
      w^2 \partial_2 \left( \frac{u}{w} \right) + 
      v^2 \partial_3 \left( \frac{w}{v} \right) = 0
\end{equation}
From (\ref{jac3dc1}) it is clear that $\{u(x),v(x),w(x)\}$ are solutions if: 
\begin{equation}
      \label{uvc1}
	\frac{v}{u}=\alpha (x_2,x_3) \;\: \Longrightarrow \;\:\;
	\left\{ \begin{array}{ccc}
	u & = & u_1(x_2,x_3) \psi_1(x) \\
	v & = & v_1(x_2,x_3) \psi_1(x)
	\end{array} \right.
\end{equation}
\begin{equation}
      \label{uwc1}
	\frac{u}{w}=\beta (x_1,x_3) \;\: \Longrightarrow \;\:\;
	\left\{ \begin{array}{ccc}
	u & = & u_2(x_1,x_3) \psi_2(x) \\
	w & = & w_2(x_1,x_3) \psi_2(x)
	\end{array} \right.
\end{equation}
\begin{equation}
      \label{vwc1}
	\frac{w}{v}=\gamma (x_1,x_2) \;\: \Longrightarrow \;\:\;
	\left\{ \begin{array}{ccc}
	v & = & v_3(x_1,x_2) \psi_3(x) \\
	w & = & w_3(x_1,x_2) \psi_3(x)
	\end{array} \right.
\end{equation}
In (\ref{uvc1}-\ref{vwc1}) the functions $\{ \alpha ,\beta , \gamma , u_i , v_i , w_i , 
\psi _j \}$, with $i \in \{ 1,2,3 \}$ and $j=1,2,3$, are $C^1(\Omega)$ and nonvanishing 
arbitrary functions of their respective arguments. A family of solutions of equations 
(\ref{uvc1}-\ref{vwc1}) is found if we assume that $\psi_j(x) \equiv \psi_j(x_j)$ 
for all $j=1,2,3$. Then, taking also into account Theorem 2.1 and Definition 2.3 we arrive 
to result (\ref{sol1expl}). \hfill Q.E.D.

\mbox{}

{\em Corollary 2.5: \/} For every open domain $\Omega \subset I \!\! R^3$, solution 
(\ref{sol1expl}) can be written as:
\begin{equation}
	\label{sol1}
	J_{ij}(x)= \eta (x) \psi_i(x_i) \psi_j(x_j) \sum_{k=1}^{3}
	\epsilon _{ijk} \phi_k(x_k)
\end{equation}
where indexes $i,j$ run from 1 to 3, $\{ \eta , \psi_i , \phi_i \}$ are arbitrary 
$C^1(\Omega)$ functions of their respective arguments which do not vanish in $\Omega$ and 
$\epsilon$ is the Levi-Civita symbol.

\mbox{}

{\em Definition 2.6: \/} For every open domain $\Omega \subset I \!\! R^3$, the subset of 
$\Gamma _{[u,v,w]}(\Omega)$ composed of those solutions of equation (\ref{jac3df}) 
characterized by Theorem 2.4 will be denoted $\Delta (\Omega)$.

\mbox{}

The family of solutions $\Delta (\Omega)$ is very general, therefore containing numerous 
previously known structure matrices of very diverse three-dimensional systems as particular 
cases, as it can be seen in detail in Table I. Of special relevance are the Lie-Poisson 
structure matrix associated to the Lie algebra so(3) (for which $\psi_i(x_i)=1$, 
$\phi_i(x_i)=x_i$ and $\eta =1$) as well as the separable matrices$^{14 \:}$ ($\phi_i(x_i)=$
constant, $\eta =1$). It is worth recalling that the time dependence of some of the structure 
matrices enumerated in Table I is immaterial in this context, since the Jacobi equations are 
time-independent and therefore time plays the only role of a parameter in the solutions. 

As anticipated in Section I, the generality of solutions (\ref{sol1}) is not an obstacle in 
what regards the characterization of their main properties. We begin by the symplectic 
structure and the Casimir invariant. 

\mbox{}

{\em Proposition 2.7: \/} For every open subset $\Omega \subset I \!\! R^3$, the rank of the 
Poisson structures belonging to $\Delta (\Omega)$ is constant in $\Omega$ and equal to 2, and 
a Casimir function of the family of solutions (\ref{sol1}) forming $\Delta (\Omega)$ is 
\begin{equation}
	\label{casim1}
	C(x) = \sum_{i=1}^{3} \int \frac{\phi_i(x_i)}{\psi_i(x_i)} \mbox{d}x_i 
\end{equation}
Moreover, the Casimir invariant (\ref{casim1}) is globally defined in $\Omega$ and 
$C^2(\Omega)$.

\mbox{}

{\em Proof: \/} The rank is constant in $\Omega$ and has value 2 as a consequence of the 
nonvanishing properties of functions $\{ \eta , \psi _i , \phi _i \}$. In addition, 
according to the Pfaffian method,$^{41 \:}$ which is the simplest in this case, the Casimir 
function is found to be the solution of the system
\[
	\sum_{i=1}^{3} \frac{\phi_i(x_i)}{\psi_i(x_i)} \mbox{d}x_i = 0 
\]
The integration is immediate and leads to (\ref{casim1}). The remaining properties of the 
Casimir invariant also arise from those of functions $\phi_i$ and $\psi_i$. \hfill Q.E.D.

\mbox{}

It is interesting to notice that $\eta(x)$ does not affect neither the symplectic structure nor 
the form of the Casimir invariant. This is to be expected from the fact that it is a common  
factor of the structure functions.

We proceed now to construct globally the Darboux canonical form. 

\mbox{}

{\bf Theorem 2.8: \/} For every three-dimensional Poisson system 
\[
	\frac{\mbox{d}x}{\mbox{d} t} = {\cal J} \cdot \nabla H
\]
defined in an open domain $\Omega \subset I \!\! R^3$ and such that ${\cal J} \in \Delta 
(\Omega)$, the Darboux canonical form is accomplished globally in $\Omega$ in the new 
coordinate system $\{z_1,z_2,z_3\}$ and the new time $\tau$, where $\{z_1,z_2,z_3\}$ is related 
to $\{x_1,x_2,x_3\}$ by the diffeomorphism globally defined in $\Omega$ 
\[
	z_1 (x_1) = \int \frac{\phi_1(x_1)}{\psi_1(x_1)} \mbox{d}x_1 \;\: , \;\:\;
	z_2 (x_2) = \int \frac{\phi_2(x_2)}{\psi_2(x_2)} \mbox{d}x_2 \;\: , \;\:\;
	z_3 (x) = \sum _{i=1}^3 \int \frac{\phi_i(x_i)}{\psi_i(x_i)} \mbox{d}x_i
\]
and the new time $\tau$ is given by the time reparametrization of the form (\ref{ntt})
\[
	\mbox{d} \tau = \eta(x(z)) \phi_1(x_1(z)) \phi_2(x_2(z)) \phi_3(x_3(z)) \mbox{d} t
\]

\mbox{}

{\em Proof: \/} We begin by noticing that the Darboux theorem is applicable to family 
(\ref{sol1}) because its members have constant rank 2 everywhere in $\Omega$, as seen in 
Proposition 2.7. This is a key necessary condition$^{2 \:}$ which is verified in the case of 
$\Delta (\Omega)$. Recall also that, after a general diffeomorphism $y = y(x)$, a given 
structure matrix ${\cal J}(x)$ is transformed into another one ${\cal J'}(y)$ according to the 
tensor rule:
\begin{equation}
	\label{jdiff}
      J'_{ij}(y) = \sum_{k,l=1}^n \frac{\partial y_i}{\partial x_k} J_{kl}(x) 
	\frac{\partial y_j}{\partial x_l}
\end{equation}
The reduction can be carried out in three steps:

 - Step 1: \/ We perform a first change of variables, which is globally diffeomorphic in 
$\Omega$:
\[
	y_i (x_i) = \int \frac{\phi_i(x_i)}{\psi_i(x_i)} \mbox{d}x_i \;\; , \;\;\:\; i = 1,2,3
\]
According to (\ref{jdiff}) we arrive to:
\begin{equation}
	\label{jdarb1}
	{\cal J'}(y) = \tilde{\eta}(y) \left( \begin{array}{ccc}
	 0 & 1 & -1 \\ -1 & 0 & 1 \\ 1 & -1 & 0 \end{array} \right)
\end{equation}
where $\tilde{\eta}(y) = \eta (x(y)) \phi_1(x_1(y_1)) \phi_2(x_2(y_2)) \phi_3(x_3(y_3))$.

- Step 2: \/ We can make use of the Casimir $C=y_1+y_2+y_3$ of ${\cal J'}$ in 
(\ref{jdarb1}) and perform a second change of variables globally diffeomorphic in $I \!\! R^3
\supset \Omega ' = y(\Omega)$:
\[
	z_1 = y_1 \;\: , \;\:\: z_2 = y_2 \;\: , \;\:\: z_3 = y_1+y_2+y_3
\]
The new structure matrix can be found by means of (\ref{jdiff}):
\[
	{\cal J''}(z) = \hat{\eta}(z) \left( \begin{array}{ccc}
	 0 & 1 & 0 \\ -1 & 0 & 0 \\ 0 & 0 & 0 \end{array} \right)
\]
where $\hat{\eta}(z) = \tilde{\eta} (y(z))$.

- Step 3: \/ Finally, we can make a reparametrization of time of the form (\ref{ntt}), 
namely $\mbox{d} \tau = \hat{\eta}(z) \mbox{d} t$, where $\tau$ is the new time and 
$\hat{\eta}(z)$ is easily seen to be nonvanishing in $\Omega '' = z(y(\Omega))$ and 
$C^1(\Omega '')$. The result is, according to Theorem 2.1 and (\ref{3dpos}-\ref{3dposntt}), a 
new Poisson system with matrix
\begin{equation}
	\label{jdarb3}
	{\cal J_D}(z) =  \left( \begin{array}{ccc} 
	0 & 1 & 0 \\ -1 & 0 & 0 \\ 0 & 0 & 0 \end{array} \right)
\end{equation}
and time $\tau$. Consequently, the structure matrix ${\cal J_D}$ in (\ref{jdarb3}) is already 
the one corresponding to the Darboux canonical form. The reduction is thus globally completed.
\hfill Q.E.D.

\mbox{}

According to the remarks made in the Introduction in connection with Darboux Theorem, it is 
worth noting an interesting corollary of Theorem 2.8, namely that all the diverse structures 
shown in Table I can actually be seen$^{1 \:}$ as the global representation of the same basic 
Poisson structure (namely the Darboux one) in different systems of coordinates. This is 
obviously a consequence of the tensor transformation rule (\ref{jdiff}). However, in the case 
of Theorem 2.8 this equivalence is demonstrated globally in $\Omega$, thus exceeding the usual 
scope of Darboux theorem. Notice also how this is founded on the fact that the rank of the 
structure matrix remains constant in $\Omega$, which is ensured by the nonvanishing conditions 
verified by the solutions. Consideration of a possible variability in the value of the rank 
would lead to additional geometric issues$^{1 \:}$  not regarded in this work for the sake 
of conciseness.

We conclude the exposition of the properties of the family of solutions (\ref{sol1}) with 
a brief discussion of its associated nonlinear superposition principles. We have the starting 
result:

\mbox{}

{\em Proposition 2.9: \/} For every open domain $\Omega \subset I \!\! R^3$, let $\{u,v,w\}$ 
and $\{u^*,v^*,w^*\}$ be two elements of $\Delta (\Omega)$. Then the set $\Delta (\Omega)$ has 
the structure of abelian group with respect to the operation of inner sum $\oplus$ given by 
\begin{equation}
	\label{sumtot}
	\begin{array}{rccl}
	\oplus : & \Delta (\Omega) \times \Delta (\Omega) & \longrightarrow & \Delta (\Omega) \\
	 & ( \{u,v,w\} , \{u^*,v^*,w^*\} ) & \longrightarrow & \{u,v,w\} \oplus \{u^*,v^*,w^*\} = 
	\{uu^*,vv^*,ww^*\}
	\end{array}
\end{equation}

\mbox{}

{\em Proof: \/} It is a consequence of the factorized form of the solutions (\ref{sol1}). 
\hfill Q.E.D.

\mbox{}

In what follows it will become evident why in this context the natural definition for 
operation (\ref{sumtot}) is that of a sum. For this, we need a previous definition:

\mbox{}

{\em Definition 2.10: \/} For every open domain $\Omega \subset I \!\! R^3$, the subset of 
$\Delta (\Omega)$ composed of solutions $\{ u, v, w \}$ such that $u(x)>0$, $v(x)>0$ and 
$w(x)>0$ for all $x \in \Omega$ will be denoted $\Delta ^+ (\Omega)$.

\mbox{}

Notice that due to the nonvanishing character of the solutions forming $\Delta (\Omega)$, the 
definition of $\Delta ^+ (\Omega)$ is not very restrictive. In fact, most examples of Table I 
belong to $\Delta ^+ (\Omega)$ or do have admissible ranges of the system parameters or 
variables for which the Poisson structure is in $\Delta ^+ (\Omega)$. It is now possible to 
establish the following:

\mbox{}

{\bf Theorem 2.11: \/} For every open domain $\Omega \subset I \!\! R^3$, let $\{u,v,w\}$ and 
$\{u^*,v^*,w^*\}$ be two elements of $\Delta ^+ (\Omega)$ and let $a$ be a real number. Then 
the set $\Delta ^+ (\Omega)$ has the structure of real vector space with respect to the two 
operations of inner sum $\oplus$ given by 
\[
	\begin{array}{rccl}
	\oplus : & \Delta ^+ (\Omega) \times \Delta ^+ (\Omega) & \longrightarrow & 
	\Delta ^+ (\Omega) \\
	 & ( \{u,v,w\} , \{u^*,v^*,w^*\} ) & \longrightarrow & \{u,v,w\} \oplus \{u^*,v^*,w^*\} = 
	\{uu^*,vv^*,ww^*\}
	\end{array}
\]
and product $\otimes$ by scalars 
\[
	\begin{array}{rccl}
	\otimes : & I\!\!R \times \Delta ^+ (\Omega) & \longrightarrow & \Delta ^+ (\Omega) \\
	 & ( a , \{u,v,w\} ) & \longrightarrow & a \otimes \{u,v,w\} = \{u^a,v^a,w^a\}
	\end{array}
\]

\mbox{}

{\em Proof: \/} It is an extension of the proof of Proposition 2.9. \hfill Q.E.D.

\mbox{}

Notice the interest of the fact that the results of the operations $\oplus$ and $\otimes$ 
belong to $\Delta ^+ (\Omega)$: According to Definition 2.10 this means that the results of 
those operations are also solutions ---of the same kind $\Delta ^+ (\Omega)$--- of equation 
(\ref{jac3df}). It is also remarkable that the nonlinear superposition principle just described 
has such a general linear algebraic structure. This is certainly infrequent in the domain of 
nonlinear PDEs.

The description of the first family of solutions is thus completed. We now proceed to examine a 
second, complementary case.

\mbox{}

\begin{flushleft}
{\bf III. SECOND FAMILY OF SOLUTIONS}
\end{flushleft}

The second family of solutions arises when we consider the case in which one of the solution 
functions $\{ u,v,w \}$ is identically zero, while the remaining two are not. 

\mbox{}

{\em Definition 3.1: \/} For every open domain $\Omega \subset I \!\! R^3$, we shall denote by 
$\Gamma_{[v,w]}(\Omega)$, $\Gamma_{[u,w]}(\Omega)$ and $\Gamma_{[u,v]}(\Omega)$ the sets of 
solutions $\{ u, v, w \}$ of equation (\ref{jac3df}) defined in $\Omega$ which are of the forms 
$\{ 0, v(x), w(x) \}$, $\{ u(x), 0, w(x) \}$ and $\{ u(x), v(x), 0 \}$, respectively, where 
$u(x)$, $v(x)$ and $w(x)$ are, when present, $C^1(\Omega)$ and nonvanishing in $\Omega$.

\mbox{}

{\bf Theorem 3.2: \/} For every open subset $\Omega \subset I \!\! R^3$, the general solutions 
of equation (\ref{jac3df}) corresponding to $\Gamma_{[v,w]}(\Omega)$, $\Gamma_{[u,w]}(\Omega)$ 
and $\Gamma_{[u,v]}(\Omega)$ are, respectively,
\begin{eqnarray}
   \Gamma_{[v,w]}(\Omega) & \Longrightarrow & \{ u= 0 , v= \eta (x) , w= \eta (x) \xi (x_1,x_2) \} 
	\label{sol2u} \\
   \Gamma_{[u,w]}(\Omega) & \Longrightarrow & \{ v= 0 , w= \eta (x) , u= \eta (x) \zeta (x_1,x_3) \} 
	\label{sol2v} \\
   \Gamma_{[u,v]}(\Omega) & \Longrightarrow & \{ w= 0 , u= \eta (x) , v= \eta (x) \chi (x_2,x_3) \} 
	\label{sol2w}
\end{eqnarray}
where functions $\{ \eta , \xi , \zeta , \chi \}$ appearing in (\ref{sol2u}-\ref{sol2w}) are 
arbitrary, $C^1(\Omega)$ with regard to their respective arguments and nonvanishing in 
$\Omega$.

\mbox{}

{\em Proof: \/} It is immediate from equation (\ref{jac3df}). \hfill Q.E.D. 

\mbox{}

Accordingly, for example in the case $u=0$ we have found structure matrices of the form
\begin{equation}
	\label{j2u}
	{\cal J} = \eta (x) \left( \begin{array}{ccc}
	0  &         0      &       -1     \\
	0  &         0      & \xi(x_1,x_2) \\
	1  & - \xi(x_1,x_2) &        0 
	\end{array} \right) 
\end{equation}
where $\eta$ and $\xi$ are $C^1(\Omega)$ and nonvanishing in $\Omega$. As it can be seen, the 
overall factor considered in Theorem 2.1 already appears explicitly in (\ref{j2u}) and needs 
not be added {\em a posteriori}.

Again, numerous well-known systems from diverse fields have Poisson structures which are 
particular cases of (\ref{sol2u}), (\ref{sol2v}) or (\ref{sol2w}), as it is displayed in 
Table II for $\Gamma _{[v,w]}(\Omega)$, in Table III for $\Gamma _{[u,w]}(\Omega)$ and in Table 
IV for $\Gamma _{[u,v]}(\Omega)$.

Following the same scheme than in the previous section, we now proceed to develop the main 
properties of the solutions just found. For the sake of conciseness this shall be done only 
for the case $\Gamma _{[v,w]}(\Omega)$, given that all the corresponding algorithms and results 
are entirely analogous for $\Gamma _{[u,w]}(\Omega)$ and $\Gamma _{[u,v]}(\Omega)$.

We shall begin with the symplectic structure and Casimir invariants. Again, the Pfaffian 
method$^{41 \:}$ seems to be the simplest one in order to characterize these properties. 
From (\ref{j2u}) the Pfaffian system to be solved is easily seen to be $\xi (x_1,x_2) 
\mbox{d}x_1 + \mbox{d}x_2 = 0$. Clearly, this system cannot be solved without some additional 
information because it is very generic. In order to circumvent this difficulty, it is worth 
introducing a definition:

\mbox{}

{\em Definition 3.3: \/} Let $\Omega \subset I \!\! R^2$ be an open domain and let $\xi : 
\Omega \longrightarrow I \!\! R$ be a $C^1(\Omega)$ function which does not vanish in $\Omega$. 
We shall say that $\xi (x_1,x_2)$ is separable in $\Omega$ if it can be written in the form 
\begin{equation}
	\label{decxi}
	\xi (x_1,x_2) = \frac{\xi _1 (x_1)}{ \xi _2(x_2)}
\end{equation}
for all $(x_1,x_2) \in \Omega$, where $\xi _1(x_1)$ and $\xi _2(x_2)$ are $C^1(\Omega)$ and do 
not vanish in $\Omega$.

\mbox{}

Now note that all specific systems found in practice (see Table II) verify the property that 
$\xi (x_1,x_2)$ is separable (notice that the only exception is the last item of Table II, but 
this is not a specific system but a generic situation which does not correspond to any 
particular vector field, and therefore it does not affect the generality of (\ref{decxi})). An 
analogous property is verified for all entries of Tables III and IV. Consequently, it seems 
well justified to conclude that, typically, $\xi$ will be separable in the form indicated in 
Definition 3.3.

\mbox{}

{\em Proposition 3.4: \/} For every open domain $\Omega \subset I \!\! R^3$, if a solution of 
the form (\ref{sol2u}) belonging to $\Gamma _{[v,w]}(\Omega)$ has a $\xi (x_1,x_2)$ which is 
separable in $\Omega$ according to (\ref{decxi}), then the rank of such Poisson structure is 
constant in $\Omega$ and has value 2, and a Casimir function of the structure is 
\begin{equation}
	\label{casim23}
	C(x_1,x_2) = \int \xi _1(x_1) \mbox{d}x_1 + \int \xi _2(x_2) \mbox{d}x_2 
\end{equation}
In addition, the Casimir invariant (\ref{casim23}) is globally defined in $\Omega$ and 
$C^2(\Omega)$.

\mbox{}

{\em Proof: \/} The rank is constant and of value 2 in $\Omega$ due to the nonvanishing 
properties of $\eta$, $\xi _1$ and $\xi _2$. Additionally, taking (\ref{decxi}) into account 
the Pfaffian system to be solved$^{41 \:}$ becomes $\xi _1(x_1) \mbox{d}x_1 + \xi _2(x_2) 
\mbox{d}x_2 = 0 $. This leads to the Casimir function immediately. The remaining properties of 
the Casimir invariant are a consequence of those of $\xi _1$ and $\xi _2$. \hfill Q.E.D.

\mbox{}

The Darboux canonical form can also be computed under similar assumptions:

\mbox{}

{\bf Theorem 3.5: \/} For every three-dimensional Poisson system 
\[
	\frac{\mbox{d}x}{\mbox{d} t} = {\cal J} \cdot \nabla H
\]
defined in an open subset $\Omega \subset I \!\! R^3$ and such that ${\cal J} \in 
\Gamma _{[v,w]} (\Omega)$ is given by (\ref{sol2u}) and $\xi (x_1,x_2)$ in (\ref{sol2u}) is 
separable in $\Omega$ according to (\ref{decxi}), the Darboux canonical form is accomplished 
globally in $\Omega$ in the new coordinate system $\{y_1,y_2,y_3\}$ and the new time $\tau$, 
where $\{y_1,y_2,y_3\}$ is related to $\{x_1,x_2,x_3\}$ by the diffeomorphism globally defined 
in $\Omega$ 
\begin{equation}
	\label{dar2u}	
	y_1 = \int \xi _1(x_1) \mbox{d}x_1 + \int \xi _2(x_2)\mbox{d}x_2  \: , \;\: 
	y_2 = x_2 \: , \;\: y_3=x_3
\end{equation}
and the new time $\tau$ is given by the time reparametrization of the form (\ref{ntt})
\[
	\mbox{d} \tau = \eta(x(y)) \frac{\xi_1(x_1(y))}{\xi_2(y_2)} \mbox{d} t
\]

\mbox{}

{\em Proof: \/} Notice first that the Darboux theorem is applicable in this case$^{2 \:}$ 
because solutions of $\Gamma _{[v,w]}(\Omega)$ of the form (\ref{j2u}-\ref{decxi}) have 
constant rank 2 everywhere in $\Omega$, as anticipated in Proposition 3.4. The reduction can be 
carried out in two steps:

- Step 1: \/ The change of variables (\ref{dar2u}), which is globally diffeomorphic in 
$\Omega$, is introduced. Notice that (\ref{dar2u}) is not the only possibility but it would be 
similar, for instance, to choose $\{ y_1 = x_1, y_2=C(x_1,x_2), y_3 = x_3 \}$. From 
(\ref{jdiff}), (\ref{j2u}), (\ref{decxi}) and (\ref{dar2u}) we are led to:
\[
	{\cal J'}(y) = \tilde{\eta} (y) \left( \begin{array}{ccc}
	0 &  0 & 0  \\
	0 &  0 & 1  \\
	0 & -1 & 0 
	\end{array} \right) 
\]
where $\tilde{\eta} (y)= \eta (x(y)) \xi (x(y)) = \eta (x(y)) \xi _1(x_1(y))/ \xi _2(y_2)$.

- Step 2: \/ A reparametrization of time of the kind (\ref{ntt}), i.e. 
$\mbox{d} \tau = \tilde{\eta}(y) \mbox{d} t$, where $\tau$ is the new time and 
$\tilde{\eta}(y)$ is clearly nonvanishing in $\Omega ' = y(\Omega)$ and $C^1( \Omega ')$. The 
resulting structure matrix is 
\begin{equation}
	\label{jdarb2u}
	{\cal J_D}(y) =  \left( \begin{array}{ccc} 
	0 & 0 & 0 \\ 0 & 0 & 1 \\ 0 & -1 & 0 \end{array} \right)
\end{equation}
Since ${\cal J_D}$ in (\ref{jdarb2u}) corresponds to the Darboux canonical form, the reduction 
has been accomplished globally. \hfill Q.E.D.

\mbox{}

To complete this section, we now consider the issue of nonlinear superposition principles. We 
have the following first result:

\mbox{}

{\em Proposition 3.6: \/} For every open domain $\Omega \subset I \!\! R^3$, let $\{0,v,w\}$ 
and $\{0,v^*,w^*\}$ be two elements of $\Gamma _{[v,w]}(\Omega)$. Then the set 
$\Gamma _{[v,w]}(\Omega)$ has the structure of abelian group with respect to the operation of 
inner sum $\oplus$ given by 
\[
	\begin{array}{rccl}
	\oplus : & \Gamma _{[v,w]}(\Omega) \times \Gamma _{[v,w]}(\Omega) & \longrightarrow & 
	\Gamma _{[v,w]}(\Omega) \\
	 & ( \{0,v,w\} , \{0,v^*,w^*\} ) & \longrightarrow & \{0,v,w\} \oplus \{0,v^*,w^*\} = 
	 \{0,vv^*,ww^*\}
	\end{array}
\]

\mbox{}

{\em Proof: \/} It is similar to that of Proposition 2.9. \hfill Q.E.D.

\mbox{}

The corresponding results for $\Gamma_{[u,w]}(\Omega)$ and $\Gamma_{[u,v]}(\Omega)$ 
are completely analogous. An additional definition is required at this stage:

\mbox{}

{\em Definition 3.7: \/} For every open domain $\Omega \subset I \!\! R^3$, the subset of 
$\Gamma _{[v,w]}(\Omega)$ composed of solutions $\{ 0,v,w \}$ such that $v(x)>0$ and $w(x)>0$ 
for all $x \in \Omega$ will be denoted $\Gamma ^+_{[v,w]}(\Omega)$.

\mbox{}

Of course, Definition 3.7 can be extended straightforwardly to characterize the sets 
$\Gamma ^+_{[u,w]}(\Omega)$ and $\Gamma ^+_{[u,v]}(\Omega)$. This is omitted for the sake of 
brevity, as usual. As it was done in the previous section, it is worth noting that the 
nonvanishing character of the functions constituting $\Gamma _{[v,w]}(\Omega)$,
$\Gamma _{[u,w]}(\Omega)$ and $\Gamma _{[u,v]}(\Omega)$ implies that the definitions of 
$\Gamma ^+_{[v,w]}(\Omega)$, $\Gamma ^+_{[u,w]}(\Omega)$ and $\Gamma ^+_{[u,v]}(\Omega)$
are not very restrictive in practice, as it can be verified in the examples lists provided in 
Tables II, III and IV, respectively. It is now possible to establish the main result:

\mbox{}

{\bf Theorem 3.8: \/} For every open domain $\Omega \subset I \!\! R^3$, let $\{0,v,w\}$ and 
$\{0,v^*,w^*\}$ be two elements of $\Gamma ^+_{[v,w]}(\Omega)$ and let $a$ be a real number. 
Then the set $\Gamma ^+_{[v,w]}(\Omega)$ has the structure of real vector space with respect to 
the two operations of inner sum $\oplus$ given by
\[
	\begin{array}{rccl}
	\oplus : & \Gamma ^+_{[v,w]}(\Omega) \times \Gamma ^+_{[v,w]}(\Omega) & \longrightarrow & 
	\Gamma ^+_{[v,w]}(\Omega) \\
	 & ( \{0,v,w\} , \{0,v^*,w^*\} ) & \longrightarrow & \{0,v,w\} \oplus \{0,v^*,w^*\} = 
	 \{0,vv^*,ww^*\}
	\end{array}
\]
and product $\otimes$ by scalars 
\[
	\begin{array}{rccl}
	\otimes : & I\!\!R \times \Gamma ^+_{[v,w]}(\Omega) & \longrightarrow & 
	\Gamma ^+_{[v,w]}(\Omega) \\
	 & ( a , \{0,v,w\} ) & \longrightarrow & a \otimes \{0,v,w\} = \{0,v^a,w^a\}
	\end{array}
\]

\mbox{}

{\em Proof: \/} It is formally analogous to that of Theorem 2.11. \hfill Q.E.D.

\mbox{}

Notice that statements similar to Theorem 3.8 can be developed in parallel for 
$\Gamma ^+_{[u,w]}(\Omega)$ and $\Gamma ^+_{[u,v]}(\Omega)$. Such results are obviously in 
correspondence with those established in Theorem 2.11 of Section II for the nonlinear 
superposition principles in $\Delta ^+(\Omega)$. Most observations made there are translatable 
into the present context in a straightforward way and are therefore disregarded.

Our analysis of the solutions of (\ref{jac3df}) can now be completed. This is the aim of the 
next section.

\mbox{}

\begin{flushleft}
{\bf IV. THIRD FAMILY OF SOLUTIONS}
\end{flushleft}

Following the previous considerations, the last possibility is to look for solutions of 
(\ref{jac3df}) such that two of the three functions $\{ u, v, w \}$ are identically zero, 
while the remaining one is not. 

\mbox{}

{\em Definition 4.1: \/} The sets of solutions $\{ u, v, w \}$ of equation (\ref{jac3df}) 
defined in an open domain $\Omega \subset I \!\! R^3$ which are of the forms 
$\{ u(x), 0, 0 \}$, $\{ 0, v(x), 0 \}$ and $\{ 0, 0, w(x) \}$, where $u(x)$, $v(x)$ and $w(x)$ 
are $C^1(\Omega)$ and nonvanishing in $\Omega$, will be denoted $\Gamma_{[u]}(\Omega)$, 
$\Gamma_{[v]}(\Omega)$ and $\Gamma_{[w]}(\Omega)$, respectively.

\mbox{}

Since all the results which are going to be examined are completely analogous for 
$\Gamma_{[u]}(\Omega)$, $\Gamma_{[v]}(\Omega)$ and $\Gamma_{[w]}(\Omega)$, we shall 
concentrate without lack of generality on the analysis of $\Gamma_{[w]}(\Omega)$.

\mbox{}

{\bf Theorem 4.2: \/} For every open domain $\Omega \subset I \!\! R^3$, the general solution 
of equation (\ref{jac3df}) corresponding to $\Gamma_{[w]}(\Omega)$ consists of the sets of 
functions of the form $\{ u=0, v=0, w(x) \}$, where $w(x)$ is an arbitrary function belonging 
to $C^1(\Omega)$ and nonvanishing in $\Omega$. Analogous results hold for $\Gamma_{[u]}(\Omega)$ 
and $\Gamma_{[v]}(\Omega)$. 

\mbox{}

{\em Proof: \/} It is immediate from equation (\ref{jac3df}). \hfill Q.E.D. 

\mbox{}

Accordingly, for example in the case of $\Gamma_{[w]}(\Omega)$ we have arrived to solutions of 
the form
\begin{equation}
	\label{j3uv}
	{\cal J}(x) = \eta (x) \left( \begin{array}{ccc}
	0 &  0 & 0  \\
	0 &  0 & 1  \\
	0 & -1 & 0 
	\end{array} \right) 
\end{equation}
with $\eta (x)$ a function $C^1(\Omega)$ and nonvanishing in $\Omega$. Notice that the 
multiplication by a global factor considered in Theorem 2.1 needs not be taken into account 
here, since it is already explicit in (\ref{j3uv}). Note also that solutions described by 
Theorem 4.2 correspond to 
structure matrices which are just time reparametrizations of the Darboux canonical form. 
Consequently, this kind of solutions is very simple and is only considered here 
for the sake of completeness: The analysis of properties such as the Casimir invariants, the 
Darboux canonical form or the existence of superposition principles becomes a straightforward 
version of those considered in Sections II and III, and can therefore be omitted. In spite of 
such simplicity, examples of Poisson structures belonging to $\Gamma_{[u]}(\Omega)$, 
$\Gamma_{[v]}(\Omega)$ or $\Gamma_{[w]}(\Omega)$ are not uncommon in the literature.$^{7 \:}$ 
In addition, it is worth mentioning that there is an important category of particular cases of 
(\ref{j3uv}) which are present in diverse systems, namely the structure matrices associated to 
the Lie algebra so(3) when expressed in certain noncartesian coordinates. The simplest 
possibility is probably that of spherical coordinates:$^{2 \:}$
\[
	{\cal J}_{so(3)}(\rho , \theta , \varphi) = - \frac{1}{ \rho \sin \varphi } 
	\left( \begin{array}{ccc} 0 & 0 & 0  \\ 0 & 0 & 1  \\ 0 & -1 & 0 \end{array} \right) 
\]
Additional instances of (\ref{j3uv}) arising from the Lie algebra so(3) for other choices of 
the coordinate system are also of customary use.$^{19 \:}$ 

\mbox{}

\begin{flushleft}
{\bf V. FINAL REMARKS}
\end{flushleft}

Some new families of skew-symmetric solutions of the three-dimensional Jacobi equations have 
been presented. They have been developed according to a systematic plan consisting in 
examining solutions of equation (\ref{jac3df}) such that (i) none of the functions 
$\{ u,v,w \}$ is identically zero (Section II); (ii) one of them is identically zero (Section 
III); (iii) two of them are identically zero (Section IV). This structuration of the 
solutions naturally embraces all nontrivial possibilities. The three resulting families have 
some remarkable properties already anticipated: 
\begin{itemize}
\item They generalize many already known Poisson structures from well-known systems, which now 
become particular cases. Therefore the new solutions unify in a common framework those 
structures, which seemed to be unrelated. Several lists of such systems are provided in Tables 
I to IV.
\item This unification allows the development of simultaneous 
methods of analysis valid for every particular system, thus avoiding a case-by-case strategy. 
Specifically, it has been shown how to construct explicitly the Casimir invariant and the 
Darboux canonical form. This is interesting, as far as the determination of the Darboux 
coordinates is a nontrivial task only known for a limited sample of systems. Moreover, in this 
work such coordinates have been charazerized globally in phase space, therefore beyond the 
usual scope of Darboux' theorem, which only ensures a local reduction. 
\item This unifying approach has uncovered the existence of nonlinear superposition principles 
within the families of solutions of (\ref{jac3df}). Such principles obey well defined linear 
algebraic structures, which is a uncommon property in the framework of nonlinear PDEs.
\end{itemize}
These results seem to indicate that the direct search of solutions of the Jacobi equations is 
a sensible line of research not only from a purely theoretical or a classification perspective, 
but also from the point of view of the analysis of Poisson structures, as well as 
mathematically interesting as an example of nonlinear system of PDEs. Of course, the 
three-dimensional scenario is probably the simplest nontrivial representative case in all those 
senses, but it is worth recalling that a similar strategy has already produced some novel 
results in the most general $n$-dimensional situation.$^{14 \:}$ Therefore, it seems justified 
to conclude that this philosophy will be a source of further advances in the understanding of 
finite-dimensional Poisson structures. 

\mbox{}

\mbox{}

\begin{flushleft}
{\bf Acknowledgements}
\end{flushleft}

I am indebted to Prof. V\'{\i}ctor Fair\'{e}n for fruitful discussions. This work was supported 
in part by the European Union (Esprit WG 24490). 

\pagebreak
\begin{flushleft}
{\bf References and notes}
\end{flushleft}
   $^1$ A. Weinstein, J. Diff. Geom. {\bf 18}, 523 (1983). \newline 
   $^2$ P. J. Olver, {\em Applications of Lie Groups to Differential Equations\/} 
	(Springer-Verlag, New York, 1993), 2nd ed. \newline 
   $^3$ K. H. Bhaskara, Proc. Indian Acad. Sci. Math. Sci. {\bf 100}, 189 (1990). \newline
   $^4$ K. H. Bhaskara and K. Rama, J. Math. Phys. {\bf 32}, 2319 (1991). \newline
   $^5$ G. B. Byrnes, F. A. Haggar and G. R. W. Quispel, Physica A {\bf 272}, 99 
	(1999). \newline
   $^6$ L. Cair\'{o} and M. R. Feix, J. Phys. A {\bf 25}, L1287 (1992). \newline 
   $^7$ D. David and D. D. Holm, J. Nonlinear Sci. {\bf 2}, 241 (1992). \newline 
   $^8$ P. Gao, Phys. Lett. A {\bf 273}, 85 (2000). \newline
   $^9$ J. Goedert, F. Haas, D. Hua, M. R. Feix and L. Cair\'{o}, J. Phys. A {\bf 27}, 6495 
	(1994). \newline 
   $^{10}$ H. G\"{u}mral and Y. Nutku, J. Math. Phys. {\bf 34}, 5691 (1993). \newline
   $^{11}$ F. Haas and J. Goedert, Phys. Lett. A {\bf 199}, 173 (1995). \newline 
   $^{12}$ B. Hern\'{a}ndez--Bermejo and V. Fair\'{e}n, Phys. Lett. A 
      {\bf 234}, 35 (1997). \newline 
   $^{13}$ B. Hern\'{a}ndez--Bermejo and V. Fair\'{e}n, J. Math. Phys. {\bf 39}, 6162 
	(1998). \newline
   $^{14}$ B. Hern\'{a}ndez--Bermejo and V. Fair\'{e}n, Phys. Lett. A {\bf 271}, 258 
	(2000). \newline 
   $^{15}$ S. A. Hojman, J. Phys. A {\bf 24}, L249 (1991). \newline 
   $^{16}$ S. A. Hojman, J. Phys. A {\bf 29}, 667 (1996). \newline 
   $^{17}$ D. D. Holm and K. B. Wolf, Physica D {\bf 51}, 189 (1991). \newline 
   $^{18}$ S. Lie, {\em Theorie der Transformationsgruppen\/} (B. G. Teubner, Leipzig, 1888). 
	\newline
   $^{19}$ R. G. Littlejohn, AIP Conf. Proc. {\bf 88}, 47 (1982). \newline 
   $^{20}$ Z.-J. Liu and P. Xu, Lett. Math. Phys. {\bf 26}, 33 (1992). \newline
   $^{21}$ C. A. Lucey and E. T. Newman, J. Math. Phys. {\bf 29}, 2430 (1988). \newline
   $^{22}$ Y. Nutku, Phys. Lett. A {\bf 145}, 27 (1990). \newline 
   $^{23}$ Y. Nutku, J. Phys. A {\bf 23}, L1145 (1990). \newline 
   $^{24}$ V. Perlick, J. Math. Phys. {\bf 33}, 599 (1992). \newline
   $^{25}$ G. Picard and T. W. Johnston, Phys. Rev. Lett. {\bf 48}, 1610 (1982). \newline
   $^{26}$ M. Plank, J. Math. Phys. {\bf 36}, 3520 (1995). \newline 
   $^{27}$ M. Plank, Nonlinearity {\bf 9}, 887 (1996). \newline
   $^{28}$ M. Plank, SIAM J. Appl. Math. {\bf 59}, 1540 (1999). \newline 
   $^{29}$ J.-L. Thiffeault and P. J. Morrison, Physica D {\bf 136}, 205 (2000). \newline
   $^{30}$ J. Gibbons, D. D. Holm and B. Kupershmidt, Phys. Lett. A 
      {\bf 90}, 281 (1982); D. D. Holm and B. A. Kupershmidt, Phys. Lett. A 
      {\bf 91}, 425 (1982); D. D. Holm and B. A. Kupershmidt, Phys. Lett. A 
      {\bf 93}, 177 (1983); J. E. Marsden, R. Montgomery, P. J. Morrison and 
      W. B. Thompson, Ann. Phys. (N.Y.) {\bf 169}, 29 (1986). \newline 
   $^{31}$ D. D. Holm, Physica D {\bf 17}, 1 (1985); D. Lewis, J. Marsden, R. Montgomery 
	and T. Ratiu, Physica D {\bf 18}, 391 (1986); P. J. Morrison, Rev. Mod. Phys. {\bf 70}, 
	467 (1998). \newline 
   $^{32}$ D. D. Holm, Phys. Lett. A {\bf 114}, 137 (1986); P. J. Morrison and J. M. Greene, 
	Phys. Rev. Lett. {\bf 45}, 790 (1980). \newline
   $^{33}$ D. D. Holm and B. A. Kupershmidt, Physica D {\bf 6}, 347 (1983); 
	J. E. Marsden and A. Weinstein, Physica D {\bf 4}, 394 (1982). \newline 
   $^{34}$ R. D. Hazeltine, D. D. Holm and P. J. Morrison, J. Plasma Phys. {\bf 34}, 103 
	(1985). \newline 
   $^{35}$ I. E. Dzyaloshinskii and G. E. Volovick, Ann. Phys. (N.Y.) {\bf 125}, 67 
	(1980). \newline 
   $^{36}$ J. R. Cary and R. G. Littlejohn, Ann. Phys. (N.Y.) {\bf 151}, 1 (1983);
	 R. G. Littlejohn, J. Math. Phys. {\bf 20}, 2445 (1979); 
       R. G. Littlejohn, J. Math. Phys. {\bf 23}, 742 (1982). \newline 
   $^{37}$ H. D. I. Abarbanel, D. D. Holm, J. E. Marsden and T. Ratiu, 
      Phys. Rev. Lett. {\bf 52}, 2352 (1984); D. D. Holm, J. E. Marsden, T. 
      Ratiu and A. Weinstein, Phys. Rep. {\bf 123}, 1 (1985); D. Lewis, J. 
      Marsden and T. Ratiu, J. Math. Phys. {\bf 28}, 2508 (1987). \newline
   $^{38}$ J. C. Simo, D. Lewis and J. E. Marsden, Arch. Rational Mec. Anal. {\bf 115}, 15 
	(1991); J. C. Simo, T. A. Posbergh and J. E. Marsden, Phys. Rep. {\bf 193}, 279 (1990); 
	J. C. Simo, T. A. Posbergh and J. E. Marsden, Arch. Rational Mec. Anal. {\bf 115}, 61 
	(1991). 
      \newline 
   $^{39}$ D. David, D. D. Holm and M. V. Tratnik, Phys. Lett. A 
      {\bf 137}, 355 (1989); D. David, D. D. Holm and M. V. Tratnik, Phys. 
      Lett. A {\bf 138}, 29 (1989); D. David, D. D. Holm and M. V. Tratnik, 
      Phys. Rep. {\bf 187}, 281 (1990). \newline 
   $^{40}$ F. Magri, J. Math. Phys. {\bf 19}, 1156 (1978); P. J. Olver, 
      Phys. Lett. A {\bf 148}, 177 (1990).  \newline 
   $^{41}$ B. Hern\'{a}ndez--Bermejo and V. Fair\'{e}n, Phys. Lett. A {\bf 241}, 148 (1998);
	T. W. Yudichak, B. Hern\'{a}ndez--Bermejo and P. J. Morrison, Phys. Lett. A {\bf 260}, 
	475 (1999). \newline
   $^{42}$ P. J. Morrison and R. D. Hazeltine, Phys. Fluids {\bf 27},  
      886 (1984). 

\pagebreak

\noindent TABLE I. Some Poisson structures reported in the literature which are particular 
cases of solution (\ref{sol1}). The original notations have been maintained for the 
parameters.

\mbox{}

\mbox{}

\begin{tabular}{llllc} \hline \hline
& \mbox{} & \mbox{} & \mbox{} & \mbox{} \\
System & Reference(s) & $\psi_i(x_i)$ & $\phi_i(x_i)$ & $\eta(x)$  \\ 
& \mbox{} & \mbox{} & \mbox{} & \mbox{} \\ \hline 
& \mbox{} & \mbox{} & \mbox{} & \mbox{} \\
Euler top & [2, pp. 397-8]  & 1 & $x_i$ & 1 \\ 
& \mbox{} & \mbox{} & \mbox{} & \mbox{} \\
& \mbox{} & \mbox{} & \mbox{} & \mbox{} \\
Kermack-McKendrick & [10, $J_1$ in Eq. (177)]; & 1 & 1 & $rx_1x_2$ \\ 
\vspace{-2mm} & \mbox{} & \mbox{} & \mbox{} & \mbox{} \\
 & [23, Eq. (5)]& & & \\ 
& \mbox{} & \mbox{} & \mbox{} & \mbox{} \\
& \mbox{} & \mbox{} & \mbox{} & \mbox{} \\
Lorenz (1) & [9, Table 3] & 1 & $\phi_1=(r/ \sigma)x_1e^{(1-\sigma) t}$ & $1/2$ \\ 
\vspace{-2mm} & \mbox{} & \mbox{} & \mbox{} & \mbox{} \\
 & & & $\phi_2 = -x_2 e^{(\sigma -1)t}$ & \\
\vspace{-2mm} & \mbox{} & \mbox{} & \mbox{} & \mbox{} \\
 & & & $\phi_3 = x_3 e^{(1-3 \sigma )t}$ & \\ 
& \mbox{} & \mbox{} & \mbox{} & \mbox{} \\
& \mbox{} & \mbox{} & \mbox{} & \mbox{} \\
Lotka-Volterra & [10, $J_1$ in Eq. (87)]; & $x_i$ & $\phi_1=-1$ & 1 \\ 
\vspace{-2mm} & \mbox{} & \mbox{} & \mbox{} & \mbox{} \\
 & [22, Eq. (18)] & & $\phi_2=-bc$& \\
\vspace{-2mm} & \mbox{} & \mbox{} & \mbox{} & \mbox{} \\
 & & & $\phi_3=c$& \\ 
& \mbox{} & \mbox{} & \mbox{} & \mbox{} \\
& \mbox{} & \mbox{} & \mbox{} & \mbox{} \\
Lotka-Volterra & [10, $J_2$ in Eq. (87)]; & $\psi_1=cx_1$ & $\phi_1=x_1$ & 1 \\ 
\vspace{-2mm} & \mbox{} & \mbox{} & \mbox{} & \mbox{} \\
 & [22, Eq. (19)] & $\psi_2=x_2$ & $\phi_2=-(x_2+\nu)$& \\
\vspace{-2mm} & \mbox{} & \mbox{} & \mbox{} & \mbox{} \\
 & & $\psi_3=x_3$ & $\phi_3=ax_3+\mu$& \\ 
& \mbox{} & \mbox{} & \mbox{} & \mbox{} \\
& \mbox{} & \mbox{} & \mbox{} & \mbox{} \\
Lotka-Volterra  & [13, Eq. (11)]; & $x_i$ & $\phi_1=K_{23}$ & 1 \\ 
\vspace{-2mm} & \mbox{} & \mbox{} & \mbox{} & \mbox{} \\
and Generalized & [26, Eq. (10)] & & $\phi_2=K_{31}$ & \\
\vspace{-2mm} & \mbox{} & \mbox{} & \mbox{} & \mbox{} \\
Lotka-Volterra  & & & $\phi_3=K_{12}$ & \\ 
& \mbox{} & \mbox{} & \mbox{} & \mbox{} \\ \hline \hline
\end{tabular}

\mbox{} \hfill {\em (CONTINUED) \hspace{6mm}}

\pagebreak

\noindent TABLE I. (Continued)

\mbox{}

\mbox{}

\begin{tabular}{llllc} \hline \hline
& \mbox{} & \mbox{} & \mbox{} & \mbox{} \\
System & Reference(s) & $\psi_i(x_i)$ & $\phi_i(x_i)$ & $\eta(x)$  \\ 
& \mbox{} & \mbox{} & \mbox{} & \mbox{} \\ \hline 
& \mbox{} & \mbox{} & \mbox{} & \mbox{} \\ 
Maxwell-Bloch \hspace{1.2cm} & [7, Case 3] & 1 & $\phi_1= \nu x_1$ & 1 \\ 
\vspace{-2mm} & \mbox{} & \mbox{} & \mbox{} & \mbox{} \\ 
 & & & $\phi_2= \mu x_2$ & \\
\vspace{-2mm} & \mbox{} & \mbox{} & \mbox{} & \mbox{} \\ 
 & & & $\phi_3= \nu + \mu x_3$ & \\ 
& \mbox{} & \mbox{} & \mbox{} & \mbox{} \\ 
& \mbox{} & \mbox{} & \mbox{} & \mbox{} \\ 
Ravinovich (1) & [9, Table 3] & 1 & $\phi_1=-x_1/4$ & 1 \\
\vspace{-2mm} & \mbox{} & \mbox{} & \mbox{} & \mbox{} \\ 
 & & & $\phi_2 = x_2/4$ & \\
\vspace{-2mm} & \mbox{} & \mbox{} & \mbox{} & \mbox{} \\ 
 & & & $\phi_3 = (x_3/2)e^{-2 \nu t}$ & \\ 
& \mbox{} & \mbox{} & \mbox{} & \mbox{} \\ 
& \mbox{} & \mbox{} & \mbox{} & \mbox{} \\ 
Ravinovich (2) & [9, Table 3] & 1 & $\phi_1=(x_1/4)e^{-\nu t}$ & 1 \\
\vspace{-2mm} & \mbox{} & \mbox{} & \mbox{} & \mbox{} \\ 
 & & & $\phi_2 = (x_2/4)e^{- \nu t}$ & \\
\vspace{-2mm} & \mbox{} & \mbox{} & \mbox{} & \mbox{} \\ 
 & & & $\phi_3 = -h/2$ & \\ 
& \mbox{} & \mbox{} & \mbox{} & \mbox{} \\ 
& \mbox{} & \mbox{} & \mbox{} & \mbox{} \\ 
Ravinovich (4) & [9, Table 3] & 1 & $\phi_1=-(x_1/2)e^{- \nu t}$ & 1 \\
\vspace{-2mm} & \mbox{} & \mbox{} & \mbox{} & \mbox{} \\ 
 & & & $\phi_2 = -(x_2/2)e^{\nu _1 t}$ & \\
\vspace{-2mm} & \mbox{} & \mbox{} & \mbox{} & \mbox{} \\ 
 & & & $\phi_3 = h e^{\nu _1 t}$ & \\ 
& \mbox{} & \mbox{} & \mbox{} & \mbox{} \\ 
& \mbox{} & \mbox{} & \mbox{} & \mbox{} \\ 
Ravinovich (5) & [9, Table 3] \hspace{1.5cm} & 1 \hspace{1.2cm} & 
$\phi_1=(x_1/2)e^{\nu _2 t}$ \hspace{0.9cm} & 1 \\
\vspace{-2mm} & \mbox{} & \mbox{} & \mbox{} & \mbox{} \\ 
 & & & $\phi_2 = (x_2/2)e^{- \nu _2 t}$ & \\
\vspace{-2mm} & \mbox{} & \mbox{} & \mbox{} & \mbox{} \\ 
 & & & $\phi_3 = -h e^{ \nu _2 t}$ & \\ 
& \mbox{} & \mbox{} & \mbox{} & \mbox{} \\ 
& \mbox{} & \mbox{} & \mbox{} & \mbox{} \\ 
RTW interaction (1) & [9, Table 3] & 1 & $\phi_1=x_1$ & 1 \\
\vspace{-2mm} & \mbox{} & \mbox{} & \mbox{} & \mbox{} \\ 
 & & & $\phi_2 = x_2$ & \\
\vspace{-2mm} & \mbox{} & \mbox{} & \mbox{} & \mbox{} \\ 
 & & & $\phi_3 = (1/2)e^{-2t}$ & \\ 
& \mbox{} & \mbox{} & \mbox{} & \mbox{} \\ \hline \hline
\end{tabular}

\mbox{} \hfill {\em (CONTINUED) \hspace{6mm}}

\pagebreak

\noindent TABLE I. (Continued)

\mbox{}

\mbox{}

\begin{tabular}{llllc} \hline \hline
& \mbox{} & \mbox{} & \mbox{} & \mbox{} \\
System & Reference(s) & $\psi_i(x_i)$ & $\phi_i(x_i)$ & $\eta(x)$  \\ 
& \mbox{} & \mbox{} & \mbox{} & \mbox{} \\ \hline 
& \mbox{} & \mbox{} & \mbox{} & \mbox{} \\ 
RTW interaction (3) \hspace{0.1cm} & [9, Table 3] \hspace{1.5cm} & 1 \hspace{1.2cm} & 
	$\phi_1=2x_1e^{-t}$ \hspace{15mm} & 1 \\
\vspace{-2mm} & \mbox{} & \mbox{} & \mbox{} & \mbox{} \\ 
 & & & $\phi_2 = 2x_2e^{-t}$ & \\
\vspace{-2mm} & \mbox{} & \mbox{} & \mbox{} & \mbox{} \\ 
 & & & $\phi_3 = e^{-t}$ & \\ 
& \mbox{} & \mbox{} & \mbox{} & \mbox{} \\ 
& \mbox{} & \mbox{} & \mbox{} & \mbox{} \\ 
RTW interaction (4) & [9, Table 3] & 1 & $\phi_1=2x_1e^{\gamma t}$ & 1 \\
\vspace{-2mm} & \mbox{} & \mbox{} & \mbox{} & \mbox{} \\ 
 & & & $\phi_2 = 2x_2e^{\gamma t}$ & \\
\vspace{-2mm} & \mbox{} & \mbox{} & \mbox{} & \mbox{} \\ 
 & & & $\phi_3 = e^{-(2+\gamma)t}$ & \\ 
& \mbox{} & \mbox{} & \mbox{} & \mbox{} \\ 
& \mbox{} & \mbox{} & \mbox{} & \mbox{} \\ 
RTW interaction (5) & [9, Table 3] & 1 & $\phi_1=\delta x_1e^{-2t}$ & 1 \\
\vspace{-2mm} & \mbox{} & \mbox{} & \mbox{} & \mbox{} \\ 
 & & & $\phi_2 = \delta x_2e^{-2t}$ & \\
\vspace{-2mm} & \mbox{} & \mbox{} & \mbox{} & \mbox{} \\ 
 & & & $\phi_3 = \delta /2$ & \\ 
& \mbox{} & \mbox{} & \mbox{} & \mbox{} \\ 
& \mbox{} & \mbox{} & \mbox{} & \mbox{} \\ 
Spin system & [19, Eq. (14)]  & 1 & $x_i$ & 1 \\ 
& \mbox{} & \mbox{} & \mbox{} & \mbox{} \\ \hline \hline
\end{tabular}

\pagebreak

\noindent TABLE II. Some Poisson structures reported in the literature which are particular 
cases of solution (\ref{sol2u}). The original notations have been maintained for the 
parameters. 

\mbox{}

\mbox{}

\begin{tabular}{llll} \hline \hline
& & & \\
System & Reference & $\xi(x_1,x_2)=w(x)/v(x)$ & $\eta(x)=v(x)$ \\ 
& & & \\ \hline 
& & & \\
Circle maps & [10, $J_1$ in Eq. (120)] & $-(x_2/x_1)^2$ & $(x_1x_3)^2$  \\ 
& & & \\
& & & \\
May-Leonard & [10, $J_1$ in Eq. (152)] & $(x_2/x_1)^{\alpha}$ & $(\alpha -1)^{-1}
x_2^{-\alpha}$  \\ 
& & & \\
& & & \\
Ravinovich (6) & [9, Table 3] & $(x_1/x_2)e^{2(\nu _2 - \nu _1)t}$ & 
$-(x_2/2)e^{(\nu _1 - 2\nu _2)t}$  \\ 
& & & \\
& & & \\
Ravinovich (7) & [9, Table 3] & $(x_1/x_2)e^{2(\nu _2 - \nu _3)t}$ & 
$(x_2/2)e^{- \nu _2t}$  \\ 
& & & \\
& & & \\
3D with a known  & [12, Eq. (18)] & $-f_2(x_1,x_2,t)/f_1(x_1,x_2,t)$ & $-f_1(x_1,x_2,t)$  \\ 
\vspace{-2mm} & & & \\
first integral  & & & \\ 
& & & \\ \hline \hline
\end{tabular}

\pagebreak

\noindent TABLE III. Some Poisson structures reported in the literature which are particular 
cases of solution (\ref{sol2v}). The original notations have been maintained for the 
parameters. 

\mbox{}

\mbox{}

\begin{tabular}{llll} \hline \hline
& & & \\
System & Reference(s) & $\zeta(x_1,x_3)=u(x)/w(x)$ & $\eta(x)=w(x)$ \\
& & & \\ \hline 
& & & \\
Kermack-McKendrick & [10, $J_2$ in Eq. (177)]; & $rx_1/a$ & $-ax_2$  \\ 
\vspace{-2mm} & & & \\
 & [23, Eq. (6)] & &  \\ 
& & & \\
& & & \\
Lorenz & [10, $J_1$ in Eq. (139)] & $-(2x_1)^{-1}$ & $-x_1/2$ \\ 
& & & \\
& & & \\
Lorenz (3) & [9, Table 3] & $-(\sigma /x_1)e^{(2 \sigma -1)t}$ & 
$-(x_1/2)e^{- \sigma t}$ \\ 
& & & \\
& & & \\
Lorenz (5) & [9, Table 3] & $-x_1^{-1}e^t$ & $-(x_1/2)e^{- t}$ \\ 
& & & \\
& & & \\
Maxwell-Bloch & [7, Case 1] & $x_1^{-1}$ & $\nu x_1$ \\ 
& & & \\
& & & \\
Maxwell-Bloch & [10, $J_2$ in Eq. (159)] & $(k/2g)x_1^{1-2g}x_3^{2k-1}$ & 
$k^{-1}x_1^g x_2^{1-g}x_3^{1-2k}$ \\ 
& & & \\
& & & \\
Two-level & [10, Eq. (165)] & $x_1/x_3$ & $x_3/(2x_1)$ \\ 
& & & \\ \hline \hline
\end{tabular}

\pagebreak

\noindent TABLE IV. Some Poisson structures reported in the literature which are particular 
cases of solution (\ref{sol2w}). The original notations have been maintained for the 
parameters.

\mbox{}

\mbox{}

\begin{tabular}{llll} \hline \hline 
& & & \\
System & Reference & $\chi(x_2,x_3)=v(x)/u(x)$ & $\eta(x)=u(x)$ \\
& & & \\ \hline 
& & & \\
Circle maps & [10, $J_2$ in Eq. (120)] \hspace{3mm} & $-(x_3/x_2)^2$ & $(x_1x_2)^2$ \\ 
& & & \\
& & & \\
Lorenz & [10, $J_2$ in Eq. (139)] & $x_2/x_3$ & $-x_3/2$ \\ 
& & & \\
& & & \\
Maxwell-Bloch \hspace{1cm} & [7, Case 2] & $x_2/x_3$ & $ \mu x_3$ \\ 
& & & \\
& & & \\
May-Leonard & [10, $J_2$ in Eq. (152)] & $(x_3/x_2)^{\alpha}$ & 
$(\alpha -1)^{-1} x_3^{- \alpha}$ \\ 
& & & \\
& & & \\
Ravinovich (3) & [9, Table 3] & $(x_2/x_3)e^{2(\nu_3 -\nu)t}$ & 
$(x_3/2)e^{- \nu_3t}$ \\ 
& & & \\ \hline \hline 
\end{tabular}

\end{document}